\documentclass[a4paper,conference]{IEEEtran}

\usepackage[T1]{fontenc}
\usepackage{graphicx}
  \graphicspath{{figures/}}
\usepackage{url}
\usepackage{tabularx}
\usepackage{booktabs}
\usepackage{mathtools}
\usepackage{amsmath}
  \allowdisplaybreaks
\usepackage{amsfonts}
\usepackage{amssymb}
\usepackage[dvipsnames]{xcolor}
\usepackage{soul}
\usepackage{tikz}
\usepackage{makecell}
\usepackage{xstring}
\usepackage{dblfloatfix}
\usepackage{inconsolata}
\usepackage{xurl}
\usepackage{flushend}

\providecommand{\keywords}[1]{
  \small
  \textbf{\textit{Keywords ---}} #1
}

\def\footnoterule{\relax%
  \kern-5pt
  \hbox to\columnwidth{\vrule width\columnwidth height 0.2pt}
  \kern5pt}
\makeatother

\makeatletter
\newcommand*\titleheader[1]{\gdef\@titleheader{#1}}
\AtBeginDocument{%
\let\st@red@title\@title
\def\@title{%
\bgroup\normalfont\normalsize\centering\@titleheader\par\egroup
\vskip0.2em\st@red@title}
}
\makeatother

\makeatletter
\renewcommand{\fnum@figure}{Figure \thefigure}
\makeatother

\titleheader{First International Air Transportation Research and Development Symposium (ATRDS2025)}

\title{Contrail, or not contrail, that is the question:\\ the ``feasibility'' of climate-optimal routing}

\newcommand{\specialcell}[2][c]{%
  \begin{small}%
  \begin{tabular}[#1]{@{}c@{}}%
 #2%
  \end{tabular}%
  \end{small}}

\author{%
Junzi Sun\IEEEauthorrefmark{1}, %
Xavier Olive\IEEEauthorrefmark{2} %
\\[1em]
\begin{tabular}{ccc}%
  \specialcell{\IEEEauthorrefmark{1}\small Faculty of Aerospace Engineering \\ Delft University of Technology\\ Delft, the Netherlands}
  \specialcell{\IEEEauthorrefmark{2}\small ONERA -- DTIS\\Universit\'e de Toulouse\\Toulouse, France} &
\end{tabular}
}

\newcommand{\coo}{CO\textsubscript{2}~}

\begin{document}

\maketitle

\thispagestyle{plain}
\pagestyle{plain}

\begin{abstract}
The environmental impact of aviation has been the subject of significant research efforts for several decades. The goal of reducing carbon emissions has reached a consensus across different stakeholders, which led to efforts to reduce route inefficiencies in air traffic management systems. Other non-carbon emissions, like contrails, have spurred a different kind of discussion. Some stakeholders have even rapidly moved into the pre-operational phase, aiming to reduce the formation of contrail by flight re-routing. Notably, recent projects from Google and Breakthrough Energy have been fast-tracking the operational perspective of contrail avoidance. 

In this paper, we want to address the practical challenges of implementing contrail-aware routing strategies. The research is built upon our past research that allows wind-optimal 4D trajectories to be generated considering any grid-like cost functions. The TOP tool, built upon the OpenAP aircraft performance model, allows us to analyze the trade-offs between fuel consumption and contrail avoidance. We also studied the impact of weather forecast uncertainty on contrail mitigation strategies, the effects of contrail mitigation on airspace capacity and network operations, and the implications of contrail reduction strategies on the aviation regulatory frameworks. 

We first reconstruct a dataset using OpenSky data, containing all flight trajectories over Europe on a day with significant contrail potential to conduct a data-driven analysis of these challenges. Then, we demonstrate the potential difficulties in implementing contrail-optimal routing in practice, especially concerning the uncertainties in weather forecasts, airspace capacity, and the responsibility for optimal routing. Overall, we argue that contrail optimal routing should be approached with caution, and it may not be as straightforward as promoted by some aviation stakeholders.
\end{abstract}


\section{Introduction}

The aviation sector contributes roughly 3\% of global \coo emissions, a figure that is likely to grow with increasing air travel demand. Transitioning to net-zero emissions in aviation is fraught with challenges: the availability of synthetic aviation fuels produced with green energy remains limited, current aircraft technology is constrained in its fuel efficiency improvements, and the economic burden of adopting new decarbonization measures is substantial. In practice, reducing emissions involves not only developing and scaling up synthetic aviation fuels but also implementing operational changes that require coordinated efforts among regulators, manufacturers, and airline operators. This intricate interplay of technical, economic, and regulatory factors underscores the complexity of achieving a genuinely low-carbon aviation future.

In addition to \coo emissions, aviation contributes to climate change through the formation of contrails, which are cloud-like streaks created by the condensation of water vapor from aircraft exhaust in the cold upper atmosphere. These contrails can evolve into cirrus clouds, which trap heat in the atmosphere and have a net warming effect on the Earth's climate. The 2022 IPCC report noted that contrail clouds account for roughly 35\% of aviation's global warming impact, which is over half the impact of the world's jet fuel. However, the exact contribution of contrails to climate change remains debatable due to the largely uncertain form of the most commonly used model \cite{lee2021contribution}.

Recent research has intensified in this area, notably supported by initiatives such as Google's Project Contrails \cite{ng2023opencontrails}. This research has focused on the automatic detection of contrails in satellite images using machine learning techniques~\cite{mccloskey2021a,chevallier2023linear}, estimating areas where contrails are likely to form, and optimizing flight trajectories to avoid contrail formation~\cite{roosenbrand2023contrail, openskyreport24}. The use of ground-based cameras has also been investigated as another source of images, an effort announced by the EUROCONTROL \cite{eurocontrol2023contrail}.

Studies have shown that contrail formation can often be avoided through modest altitude adjustments, typically within 2,000 feet, while keeping fuel penalties minimal~\cite{schumann2011potential, teoh2020beyond, roosenbrand2023contrail}. The operational feasibility depends primarily on air traffic management capacity constraints and airline policies regarding cost trade-offs.
Beyond altitude adjustments, researchers have investigated contrail-optimal routing strategies that incorporate meteorological forecasts and real-time atmospheric data. By leveraging ensemble weather prediction models and optimization algorithms, flights can be re-routed to avoid regions where persistent contrails are likely to form~\cite{matthes2020climate, simorgh2024robust}.

While these contrail mitigation strategies have demonstrated potential for reducing climate impact, they introduce operational challenges, including increased fuel consumption, longer flight times, and potential congestion in contrail-free sectors that are managed by limited air traffic control resources. Recent studies have conducted cost-benefit analyses to identify optimal mitigation approaches that balance emissions reduction with economic viability~\cite{chen2012tradeoff, borella2024importance}.
The feasibility of climate-optimal routing has been proposed by some previous studies or tested in less complex airspace environments, such as trans-Atlantic routes~\cite{grewe2017feasibility, molloy2022design}, and through limited trials conducted by EUROCONTROL~\cite{eurocontrol2021reducing}. However, the practical challenges of implementing contrail-aware routing strategies remain largely unexplored. 


Optimizing individual flights to minimize contrail formation presents several challenges. Firstly, adjusting each flight's trajectory to avoid regions prone to contrail formation can incur additional environmental costs. It may result in increased \coo emissions due to longer routes or greater fuel consumption from flying at lower altitudes. Airlines also face elevated operational expenses when routing through airspace and higher operational costs. Secondly, optimizing flights on an individual basis might not yield the most efficient outcomes for the global air traffic system. These trajectory adjustments may lead to airspace capacity challenges. Thirdly, reliance on forecast data further introduces uncertainties; current weather forecasts are often insufficiently accurate for critical conditions such as humidity that influence contrail formation. Finally, the question arises as to who should be responsible for optimizing contrail reduction: airlines, who are incentivized to minimize their emissions, or air navigation service providers, who are responsible for managing airspace and air traffic flow.

In this paper, we analyze one day, with conditions favorable to contrail formation, of flight data over Europe to address these challenges and provide insights into integrating contrail reduction with global aviation climate impact optimization. Using this dataset, we explore the trade-offs between fuel consumption and contrail avoidance, assess the reliability of contrail reduction strategies under weather forecast uncertainty, examine the potential effects of monetizing contrail production, and discuss whether contrail mitigation should be implemented at the airline or network level.

The structure of the paper is organized as follows. Section \ref{sec:questions} outlines the key questions addressed in this study. Section~\ref{sec:scenario} describes the dataset and scenario used in this study. Section~\ref{sec:q1} addresses the trade-off between fuel consumption and contrail avoidance. Section~\ref{sec:q2} investigates the impact of weather forecast uncertainty on contrail reduction strategies. Section~\ref{sec:q3} explores the effects of contrail mitigation on airspace capacity and network operations. Section~\ref{sec:q4} discusses the implications of contrail reduction strategies on the aviation industry and regulatory frameworks. Finally, we present concluding remarks about the results and their implications for future research.

\section{Overview of the questions} \label{sec:questions}

In the following, we conduct a data-driven analysis to address these questions, leveraging real-world air traffic and meteorological data to quantify both the challenges and potential benefits of contrail-aware routing. We expose in this section the challenges attached to those questions:

\begin{enumerate}
\item \textbf{The trade-off between contrail avoidance and increased fuel consumption}\\
The first key question concerns the trade-off between increased fuel consumption and contrail avoidance. To what extent is it justifiable to accept a rise in \coo emissions in order to mitigate the climate impact of contrails? A crucial aspect is quantifying the amount of contrail reduction achieved for a given increase in fuel burn and determining whether the avoidance can provide climate or network benefit. This question is particularly relevant for airlines seeking to balance environmental sustainability with operational efficiency. By analyzing the relationship between fuel consumption and contrail formation, we can assess the feasibility of contrail-aware routing and its implications for airline operations.

\item \textbf{The uncertainty in weather forecast}\\
Another fundamental issue is the role of uncertainty in weather forecasts and its effect on the effectiveness of contrail mitigation efforts. To what extent can contrail-aware routing strategies be robust to forecast inaccuracies, and how do these uncertainties impact the overall climate benefits of contrail reduction in reality?
Contrail formation is highly dependent on atmospheric conditions such as temperature and humidity, which can be challenging to predict with precision. The accuracy of meteorological models in identifying contrail-prone regions is, therefore, a critical factor in determining whether contrail avoidance can be planned.

\item \textbf{The impact on airspace capacity}\\
The impact of contrail mitigation on airspace capacity and network operations is another essential consideration. What are the implications of contrail-aware routing on air traffic management efficiency, sector workload, and delay propagation? 
Implementing contrail-aware routing within existing air traffic management frameworks introduces operational constraints, particularly in congested airspace, where flexibility is already limited. Understanding the effects of contrail mitigation on airspace efficiency, sector workload, and delay propagation is crucial for assessing its feasibility. An important question is whether mitigation efforts can be seamlessly integrated into trajectory optimization frameworks without significantly disrupting air traffic flows or overwhelming air traffic controllers.

\item \textbf{The responsibility of contrail optimal routing}\\
A key question is who should be responsible for contrail mitigation efforts: airlines or air navigation service providers? Should contrail-aware routing be managed at the airline level or as a coordinated global initiative?
Introducing a \textit{contrail charge} could incentivize airlines to minimize contrail formation by incorporating it into their cost index, making mitigation an individual optimization problem similar to fuel efficiency. However, unlike \coo emissions, contrails have non-linear aggregation effects, meaning airline-level decisions may not lead to the best overall expected climate outcome.

A network-wide approach, managed by regulatory bodies or air navigation service providers, could optimize contrail reduction more effectively by considering interactions between flights and airspace constraints. However, this raises the challenge of quantifying contrail impact and designing a fair economic mechanism. Establishing a standardized metric for contrail charges or incentives would require improved atmospheric models, real-time data, and a regulatory framework to ensure compliance and effectiveness.

\end{enumerate}

We will attempt to bring elements to answer some questions based on real data. However, those elements related to policy-macking are hard to answer concretely. As regulation under uncertainties is a complex topic, we will only provide some insights and discuss the implications of our findings for future research and policy-making. 

\section{Dataset, scenario, and methodology}
\label{sec:scenario}

To address the questions outlined in the previous section, we use a dataset of flight trajectories over Europe on 20 February 2022. 

\begin{figure}[!hbtp]
  \centering
  \includegraphics[width=\columnwidth]{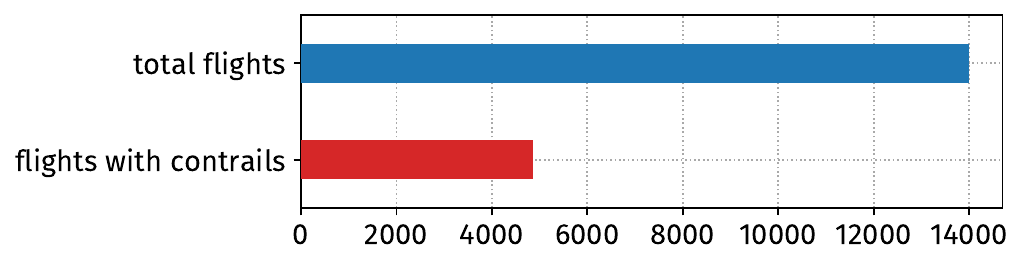}
  \caption{The number of flights (around 14k) in the dataset and flights with persistent contrails (around 5k) over Europe on 20 February 2022.}
  \label{fig:flights_vs_contrails}
\end{figure}

The date of 20 February 2022 is chosen as it represents a day when the meteorological conditions are prone to contrail formation, with high humidity levels for a large region over Western Europe. The statistics are shown in Figure \ref{fig:flights_vs_contrails}. 

\begin{figure*}[b!]
  \centering
  \includegraphics[width=\textwidth]{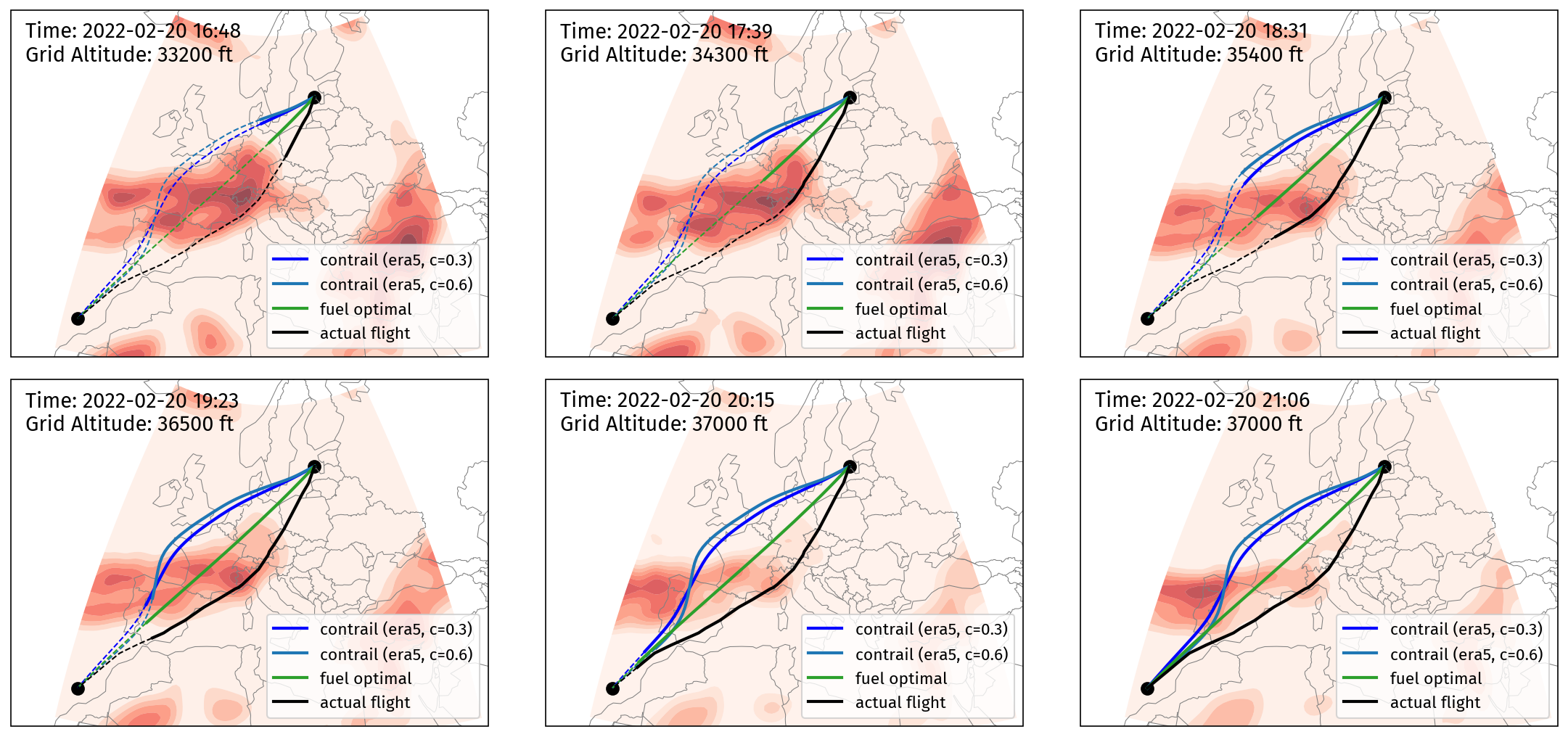}
  \caption{Example of optimal trajectories between Riga, Latvia, and the Canary Islands under different objectives. The background shading indicates contrail sensitivity: darker red denotes regions with a higher likelihood of persistent contrail formation (value 1), while lighter shades indicate lower or no contrail formation. The color gradient results from smoothing the cost grid.}
  \label{fig:optimization_example}
\end{figure*}

This single-day scenario allows us to analyze the impact of contrail-aware routing strategies on a day with significant contrail potential. We filter the dataset to include only cruise flights with altitudes above 25,000 feet with a minimum duration of 30 minutes.

We obtain the flight data from the OpenSky Network\cite{schafer2014bringing}, a global open-source platform that provides real-time air traffic data. In addition to the flight trajectories, we also use meteorological data from the ERA5 reanalysis dataset~\cite{soci2024era5} and the Meteo France ARPEGE forecast model \cite{deque1994arpege} to estimate atmospheric conditions such as temperature, humidity, and wind speed.

ARPEGE provides forecast data every 6 hours that includes a forecast of several days. We use the first 6 hours of the forecast data for each update cycle. This choice is based on our previous research, which has shown that the difference is minimal between different pre-tactical time horizons \cite{roosenbrand2024quantifying}.

We identify flights with persistent contrails by analyzing the flight trajectories and meteorological conditions. Segments of persistent contrails are calculated using the \texttt{fastmeteo} library proposed in \cite{sun2023fast}. The persistent contrails are determined to be flight trajectories satisfying both the Schmidt-Applemann Criterion and the ice-supersaturated regions. 

In total, over the approximate 14,000 flights in the dataset, around 5,000 flights are identified as having persistent contrails (Figure~\ref{fig:flights_vs_contrails}), which provides the base for the optimization and re-routing. It is worth noting that the percentage of flights with contrails is related to the meteorological conditions, and the number of flights with contrails can vary significantly from day to day.

For each flight with estimated persistent contrails, we run several alternative optimization scenarios to analyze the trade-offs between fuel consumption and contrail avoidance and the impact of weather forecast uncertainty. These optimal trajectories include:
\begin{itemize}
    \item \textbf{Fuel-optimal trajectories}: The 4D flight trajectory that minimizes fuel consumption considering the 4D wind condition from the ERA5 grids.
    \item \textbf{Contrail-optimal trajectories with ERA5 reanalysis data}: The 4D flight trajectory that minimizes contrail formation with ERA5 reanalysis data, considering the 4D wind condition from the ERA5 grids.
    \item \textbf{Contrail-optimal trajectories with ARPEGE forecast data}: The 4D flight trajectory minimizes contrail formation using ARPEGE forecast data, considering the 4D wind condition from the ARPEGE grids.
\end{itemize}

In total, five alternative trajectories are generated for each flight with persistent contrails, allowing us to compare the trade-offs between fuel consumption and contrail avoidance under different scenarios. The optimization is conducted using the TOP trajectory optimization framework \cite{sun2022top}, which has been improved to be able to integrate any four-dimensional cost grid with the non-linear optimal control optimization approach.\footnote{\url{https://openap.dev/optimize/costgrid.html}}

For the trajectories that are optimized using ERA5 and ARPEGE grids, we employ a compound objective function that combines the contrail formation cost and the fuel consumption cost, which is proposed in \cite{roosenbrand2024flight}. The formulation of the objective function is as follows:
\begin{equation}
 J = c \cdot J_\text{contrail} + (1 - c) \cdot J_\text{fuel}
\end{equation}

\noindent where $J_\text{contrail}$ is the contrail formation cost. When there is persistent contrail formation, the cost is set to 1, and when there is no persistent contrail formation, the cost is set to 0. This cost is calculated, smoothed, and provided as a four-dimensional cost grid at a resolution of 0.5 degrees. $J_\text{fuel}$ is the fuel consumption in the unit of $kg/s$, and $c$ is a weighting factor that determines the trade-off between the two objectives. In this study, we set $c$ to two different values, which are 0.3 and 0.6. When $c$ is 0.3, the optimal trajectory is more fuel efficient, while when $c$ is 0.6, the optimal trajectory is more contrail aware.

Unlike our previous research, where the fixed mass was assumed, to generate the optimized trajectories, we use a simplified mass estimation model based on the range and cruise altitude of the flight, which is proposed in a recent study \cite{tassanbi2025open}. We use the simplified two-parameter mass estimation model for A320, which can be described as follows:
\begin{equation}
  \begin{aligned}
 m &= 1.8533 \cdot d_\mathrm{km} - 1.99133 \cdot h_\mathrm{ft, cr} + 133497 - 2000 \\
 m &= \mathrm{min} (m, m_\mathrm{mtow})
  \end{aligned}        
\end{equation}

\noindent where $m$ is the aircraft mass in kg, $d_{km}$ is the flight distance in kilometers, and $h_{ft, cr}$ is the median cruise altitude in feet. The 2000 kg is subtracted to account for the amount of fuel used before reaching the cruise altitude. We also limit the mass to the maximum take-off weight $m_\mathrm{mtow}$ of the aircraft.

The fuel estimation used in this study is based on the OpenAP model \cite{sun2020openap}. Flight status, including altitude, speed, and vertical rate, from the trajectory data are used to estimate the fuel consumption of the original flight trajectory and the optimized trajectories.

In this paper, we emphasize the operational feasibility of contrail-aware routing strategies, and hence, the optimal trajectories are generated under altitude constraints. Namely, in addition to aircraft performance constraints, we are constraining the optimal cruise altitude to be within the range of 2000 feet below the initial cruise altitude and 2000 feet above the maximum cruise altitude.

\setcounter{section}{0}  
\renewcommand{\thesection}{Q\arabic{section}} 

\section{Can we reduce contrail formation with realistic excess fuel consumption?}
\label{sec:q1}

To address the first question, we analyze the trade-off between fuel consumption and contrail avoidance. We compare the fuel-optimal trajectory with contrail-optimal trajectories with coefficients $c$ of 0.3 and 0.6. The optimal trajectories are generated based on the persistent contrail conditions obtained from the ERA5 reanalysis data, which provides a more accurate representation of atmospheric conditions compared to forecast data. In this case, we want to see how much contrail reduction can be achieved with a realistic amount of extra fuel consumption.

\subsection{Example flight}

\begin{figure}[!htbp]
  \centering
  \includegraphics[width=0.95\columnwidth]{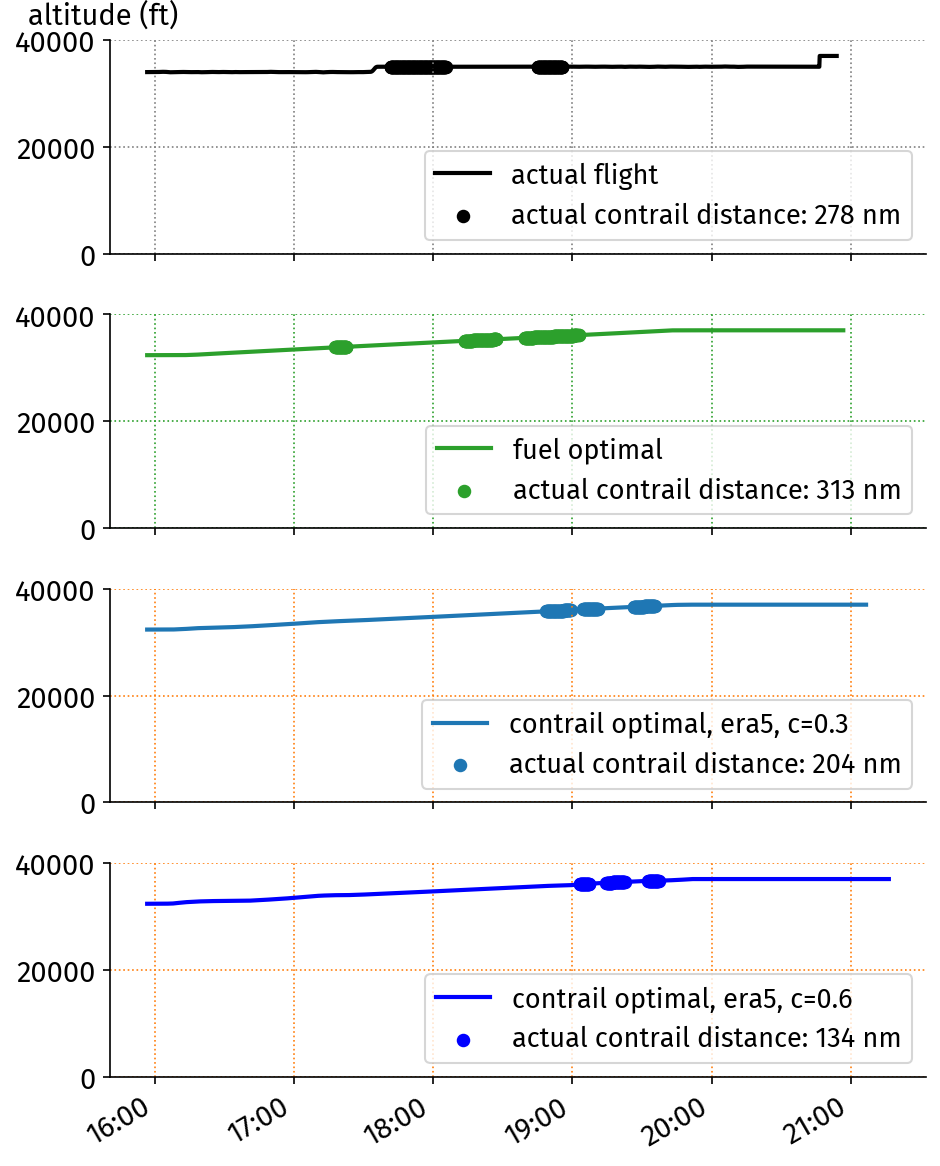}
  \caption{The altitude profiles of the example set of trajectories with different objectives and constraints.}
  \label{fig:optimization_example_altitude}
\end{figure}

In Figure~\ref{fig:optimization_example}, we show an example of a set of optimal trajectories with different objectives and constraints. This flight was chosen as it is one of the longest flights in our dataset. The trajectories are color-coded based on their objectives, which are fuel optimal trajectory, contrail optimal trajectory with $c=0.3$, and contrail optimal trajectory with $c=0.6$. 
The altitude profiles of the trajectories are shown in Figure~\ref{fig:optimization_example_altitude}, where we can observe the altitude changes made to avoid contrail formation.

In each of the sub-figures of Figure~\ref{fig:optimization_example}, we can see the time, completed flight trajectories (in solid lines), and the corresponding contrail formation regions at the closest altitude. The fuel-optimal route is generated considering the 4D wind condition from the meteorological data grids. The difference in the flight paths between the two contrail-optimal trajectories is evident. In this case, the contrail-optimal trajectories consume little additional fuel and are thereby able to avoid the more significant contrail formation regions. 

In the corresponding altitude profiles from Figure~\ref{fig:optimization_example_altitude}, we marked the exact times when persistent contrails are formed for all different trajectories. Compared to the fuel optimal trajectory, we see a significant decrease in contrail formation when extra fuel is consumed. When more fuel is consumed ($c=0.6$), the contrail formation is further reduced, which corresponds to a larger deviation from the fuel-optimal trajectory. 

It is worth noting that the optimal trajectory is strictly constrained to be within 2,000 feet below the initial altitude and 2,000 feet above the maximum cruise altitude. Under this constraint, it is not always possible to avoid contrail formation altogether, especially when the ice-supersaturated region is large vertically. Hence, the more contrail-aware trajectories tend to include larger detours, which results in a longer flight time and more fuel consumption.

\subsection{Quantitative analysis for all flights}

\begin{figure}[!ht]
  \centering
  \includegraphics[width=\columnwidth]{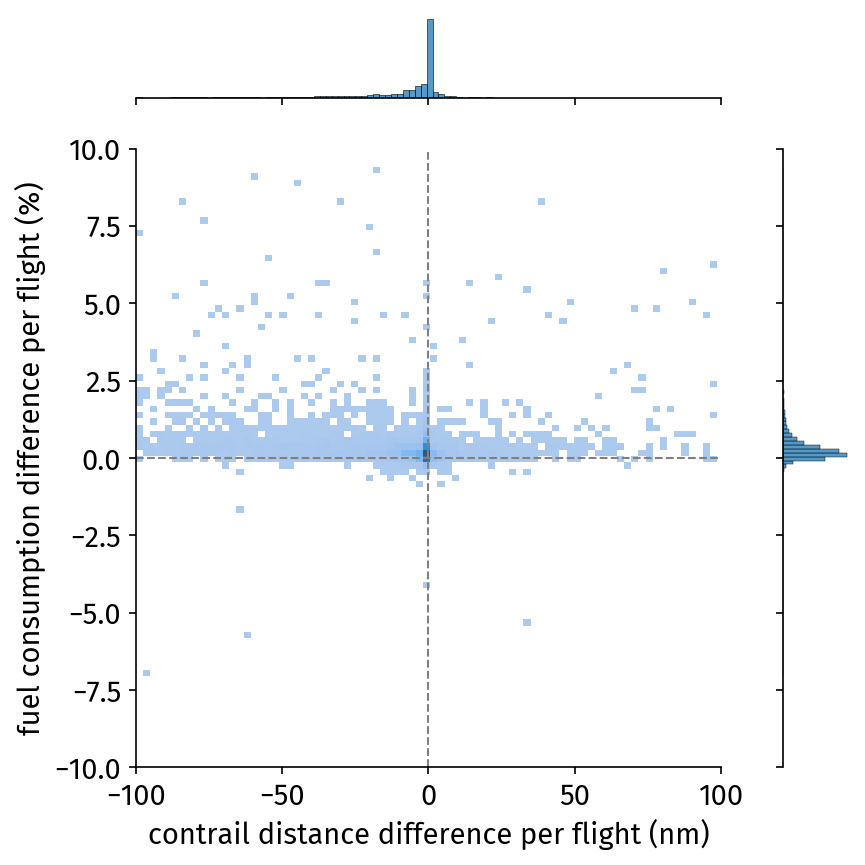}
  \caption{Changes in contrail distance with respect to fuel consumption for contrail-optimal trajectories with $c$ of 0.3}
  \label{fig:contrail_distance_vs_fuel_03}
\end{figure}

\begin{figure}[!ht]
  \centering
  \includegraphics[width=\columnwidth]{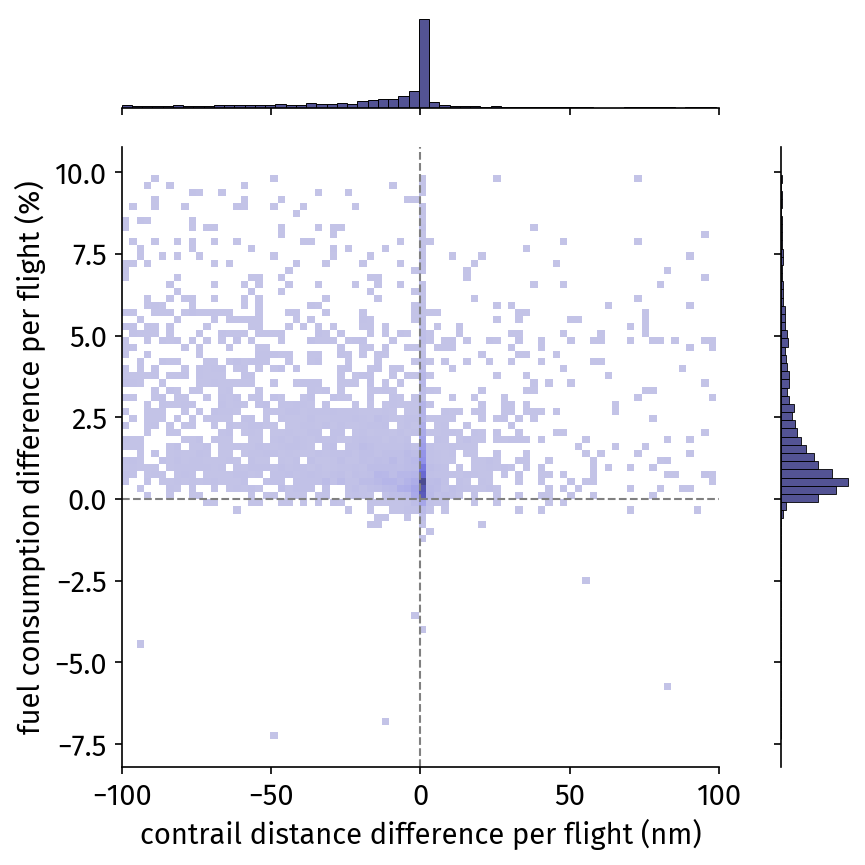}
  \caption{Changes in contrail distance with respect to fuel consumption for contrail-optimal trajectories with $c$ of 0.6}
  \label{fig:contrail_distance_vs_fuel_06}
\end{figure}

In Figure \ref{fig:contrail_distance_vs_fuel_03} and Figure \ref{fig:contrail_distance_vs_fuel_06}, we show the statistical distributions of all the flights in our dataset, together with the joint distribution of contrail distance and fuel consumption for the contrail-optimal trajectories with $c$ as 0.3 and 0.6.

The top left quadrant of the figures represents the flights where contrail formation is reduced with additional fuel consumption. The top right quadrants represent the cases where both contrail formation and fuel consumption are increased, indicating likely local minima for the optimization. The bottom quadrants represent the cases where trajectories yield less fuel consumption than the fuel-optimal trajectory, which indicates infeasible solutions due to the constraints. We can see that the majority of the flights are in the top left quadrant, indicating that contrail formation can be reduced with additional fuel consumption. However, there are also a number of flights that can not be solved within the constraints and given cost grid.

Most notably, both figures demonstrate a limited reduction in contrail formation despite increased fuel consumption, as shown by the histogram on the x-axis representing changes in contrail distance. The result suggests that contrail-aware routing strategies may be ineffective in reducing contrail formation under strict altitude constraints, even when allowing for excess fuel consumption. 

\begin{figure}[!htbp]
  \centering
  \includegraphics[width=\columnwidth]{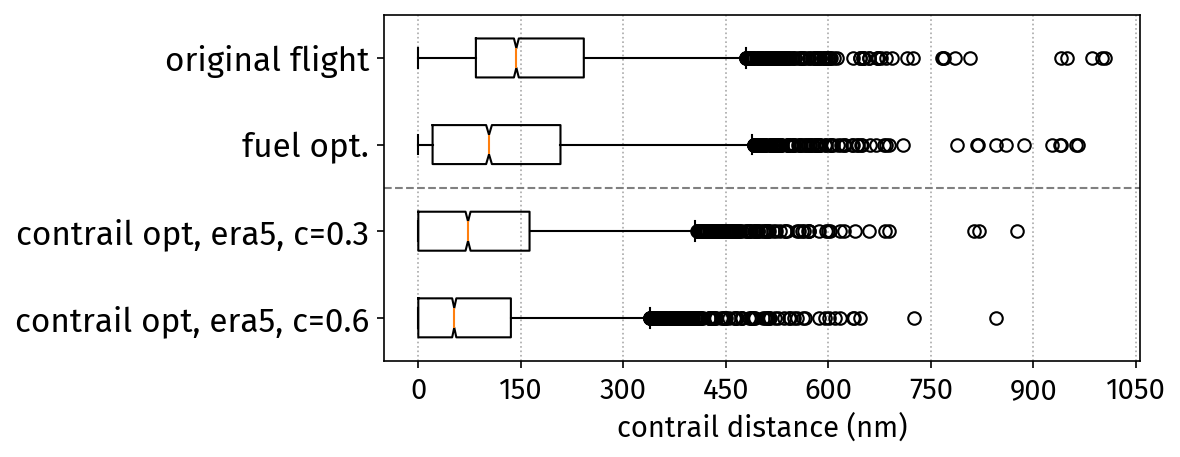}
  \caption{Contrail distance for original flights, fuel-optimal flights, and contrail-optimal trajectories with $c$ of 0.3 and 0.6}
  \label{fig:contrail_distance_era5_boxplot}
\end{figure}

Figure~\ref{fig:contrail_distance_era5_boxplot} presents a boxplot comparing contrail distances for trajectories optimized with $c$ of 0.3 and 0.6. Overall, the results indicate that contrail distances are not substantially reduced with additional fuel consumption, suggesting that the trade-off between fuel use and contrail avoidance may be unfavorable in many practical scenarios.

\section{Can we rely on weather forecast data for contrail re-routing?}
\label{sec:q2}

Many of the previous studies have been relying on the reanalysis data to estimate the contrail formation regions. However, in practice, airlines and air traffic management can only rely on weather forecast data to optimize flight trajectories during the flight planning phase. In this section, we investigate the impact of weather forecast uncertainty on contrail-aware routing strategies. Specifically, we compare the contrail-optimal trajectories generated using reanalysis data with those generated using forecast data.

\subsection{Comparison between ERA5 and ARPEGE data}

\begin{figure}[!htbp]
  \centering
  \includegraphics[width=0.49\columnwidth]{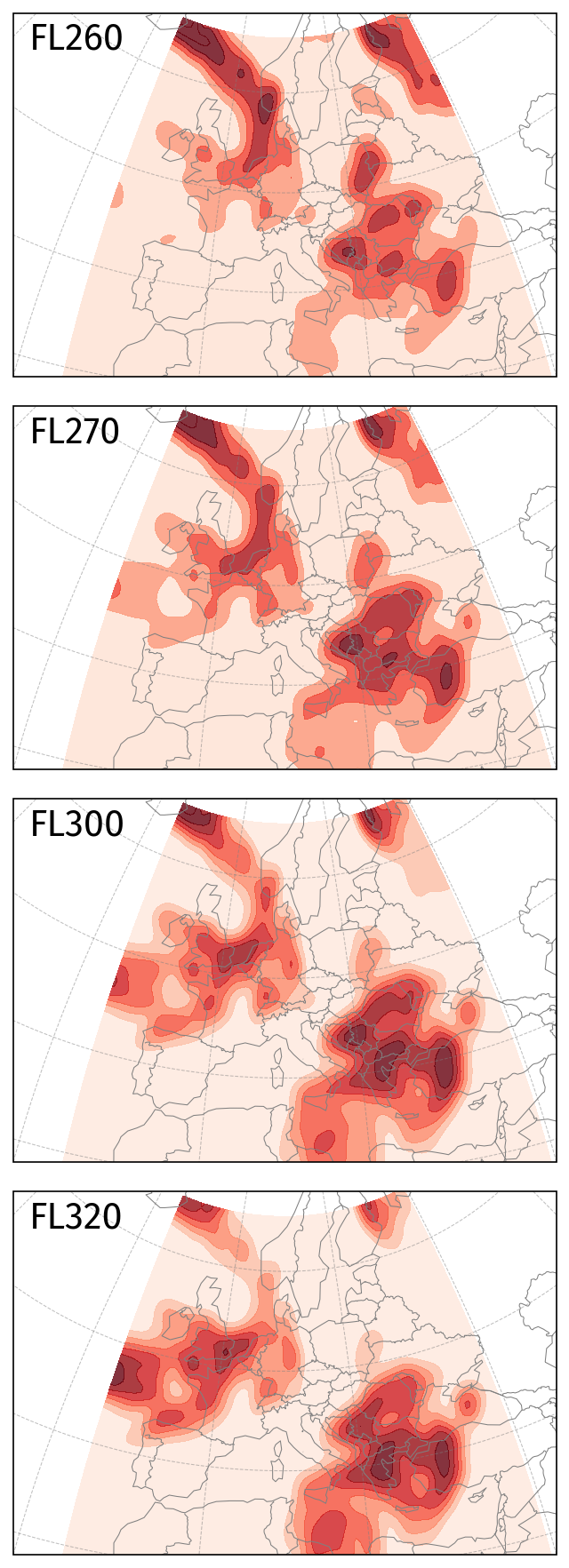}
  \includegraphics[width=0.49\columnwidth]{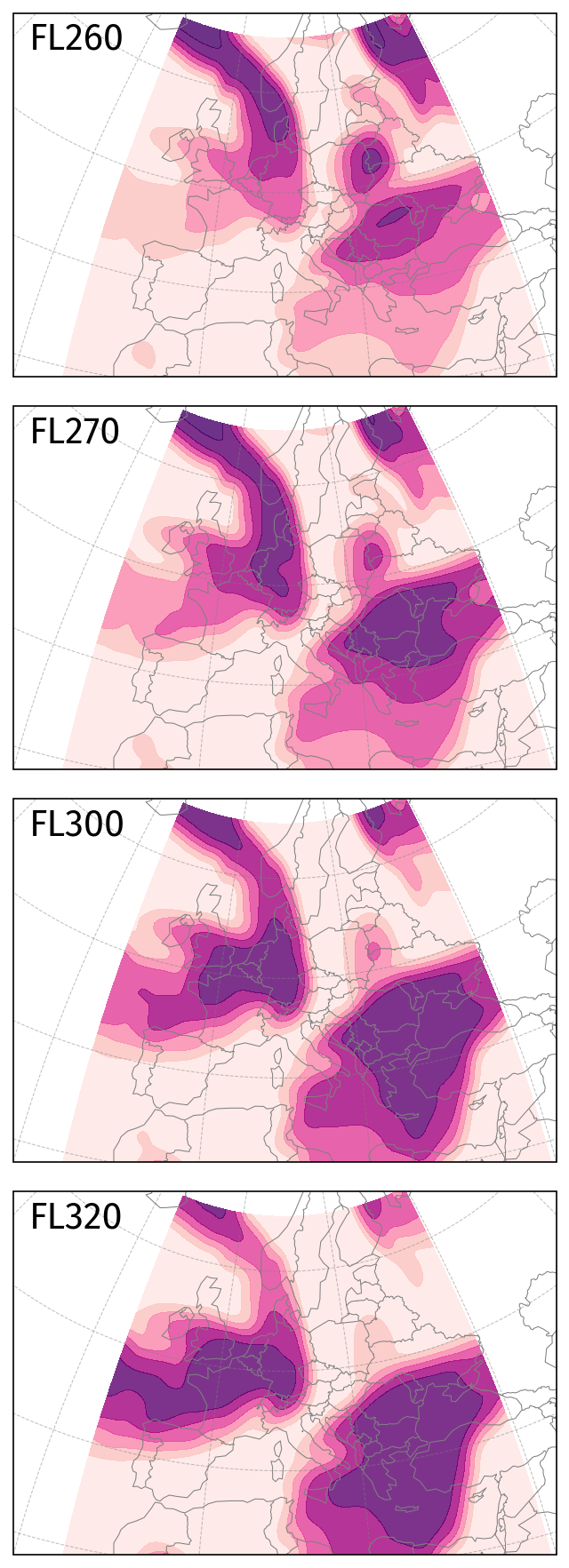}
  \caption{Persistent contrail regions obtained from ERA5 reanalysis data (left) and ARPEGE forecast data (right) over Europe on 20 February 2022. Both grids are at a resolution of 0.5 degrees. The darkest color represents the persistent contrail formation regions at the closest altitude with a numerical value of 1, and the lightest color represents the regions with no contrail formation. The color gradient is due to the smoothing of the cost grid}
  \label{fig:era5_cost_grid}
\end{figure}

Figure \ref{fig:era5_cost_grid} shows the persistent contrail formation regions over Europe on 20 February 2022, computed using both ERA5 reanalysis and ARPEGE forecast data. While both datasets identify similar regions for contrail formation, they differ in spatial resolution and precision. These differences significantly influence the resulting optimal trajectories. To improve the data quality, we applied a 3D Gaussian smoothing filter (with a standard deviation of 2 grid points) to reduce noise and sharp discontinuities in both datasets.

\begin{figure}[!htbp]
  \centering
  \includegraphics[width=0.49\columnwidth]{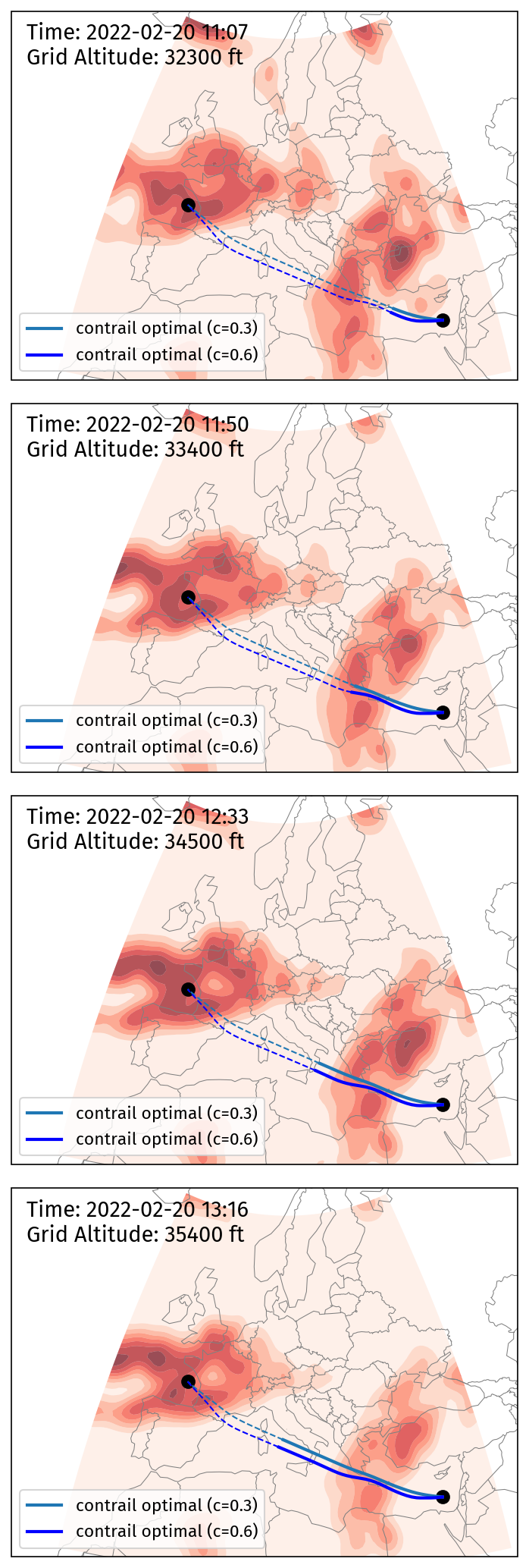}
  \includegraphics[width=0.49\columnwidth]{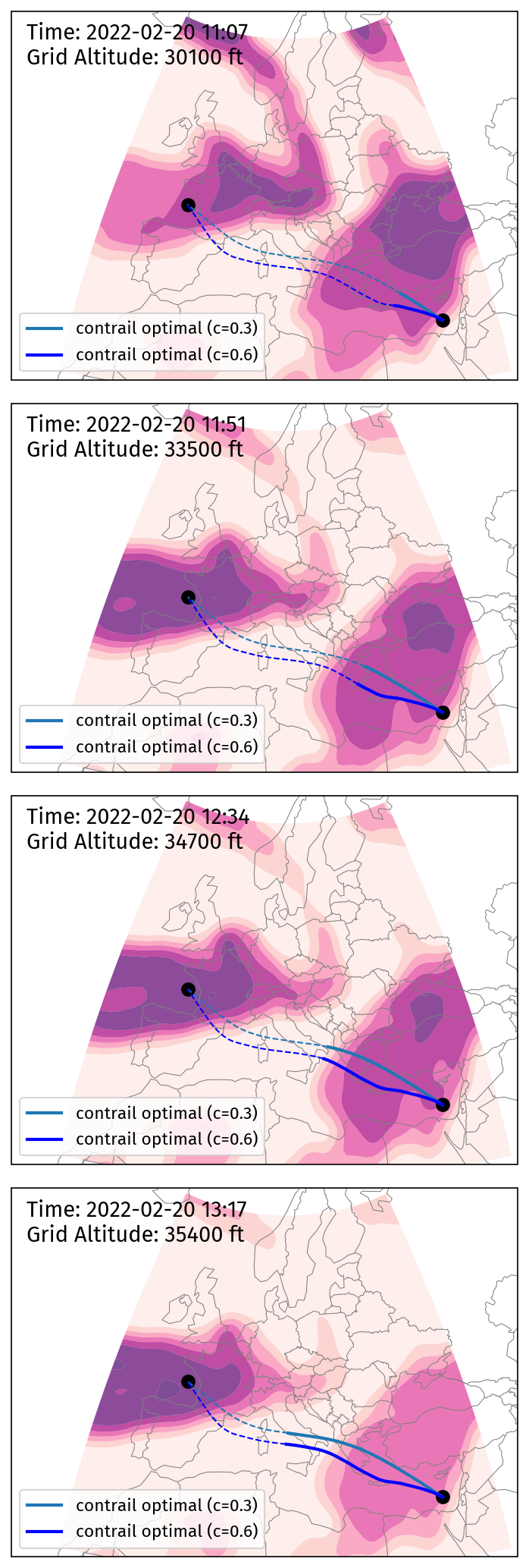}
  \caption{Contrail optimal trajectories obtained from ERA5 reanalysis data (left) and ARPEGE forecast data (right) over Europe on 20 February 2022.}
  \label{fig:optimization_example_era5_vs_arpege}
\end{figure}

\begin{figure}[!htpb]
  \centering
  \includegraphics[width=0.95\columnwidth]{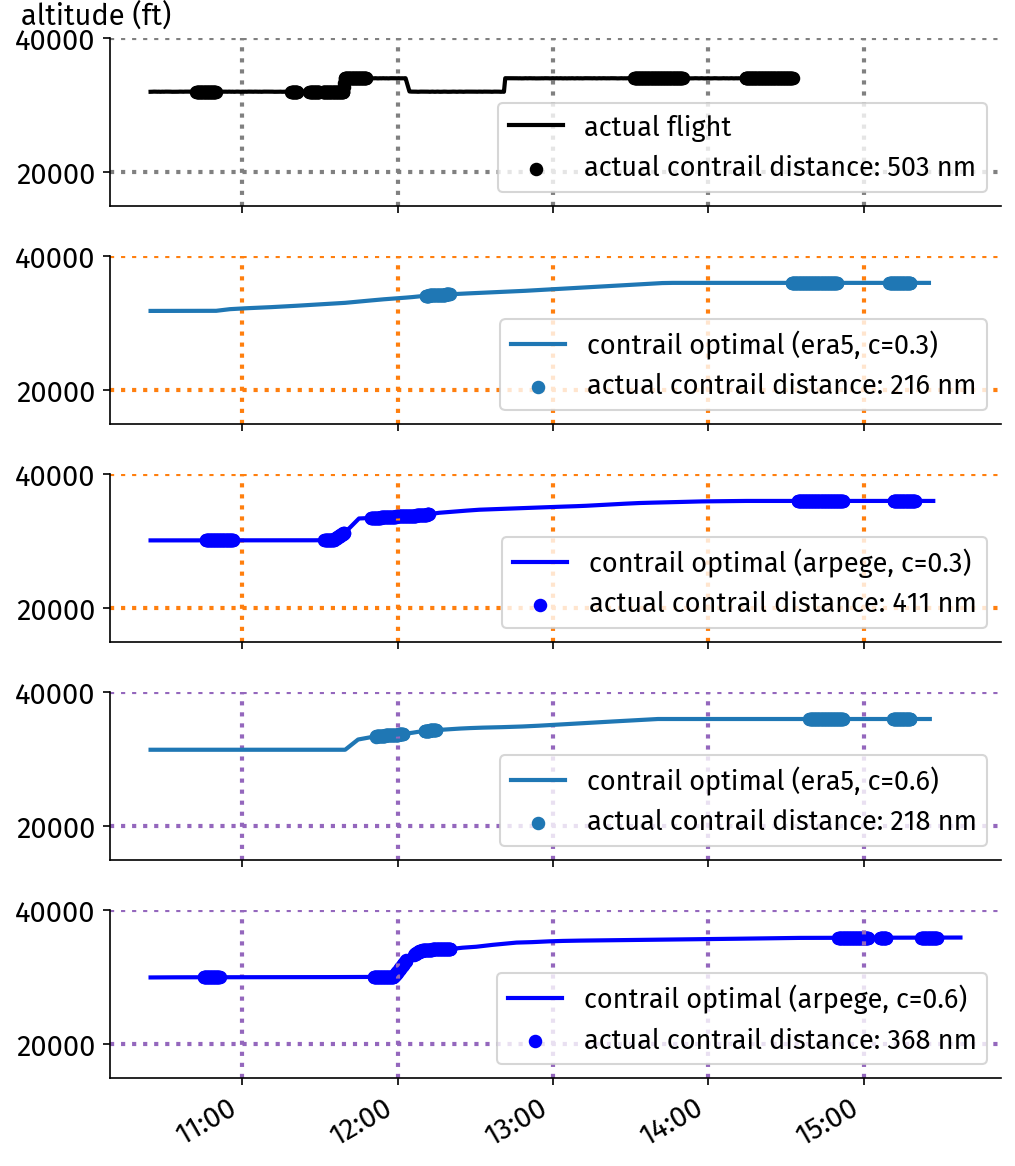}
  \caption{The altitude profiles of the example optimal trajectories conducted with ERA5 reanalysis data and ARPEGE forecast data.}
  \label{fig:optimization_example_altitude_era5_vs_arpege}
\end{figure}

To illustrate the differences between optimal trajectories generated from different data sources, Figure \ref{fig:optimization_example_era5_vs_arpege} compares contrail-optimal trajectories obtained using ERA5 reanalysis data versus ARPEGE forecast data for the same flight. Figure \ref{fig:optimization_example_altitude_era5_vs_arpege} shows the corresponding altitude profiles, with segments where contrails formed according to reanalysis data highlighted. Note that the contrail formation regions are all calculated using the ERA5 reanalysis data, which is used as the ground truth for contrail formation.

In this simple trajectory example, there is a significant discrepancy between the trajectories optimized using forecast versus reanalysis data, highlighting how weather forecast uncertainty impacts the effectiveness of contrail avoidance strategies.

\begin{figure}[!htpb]
  \centering
  \includegraphics[width=\columnwidth]{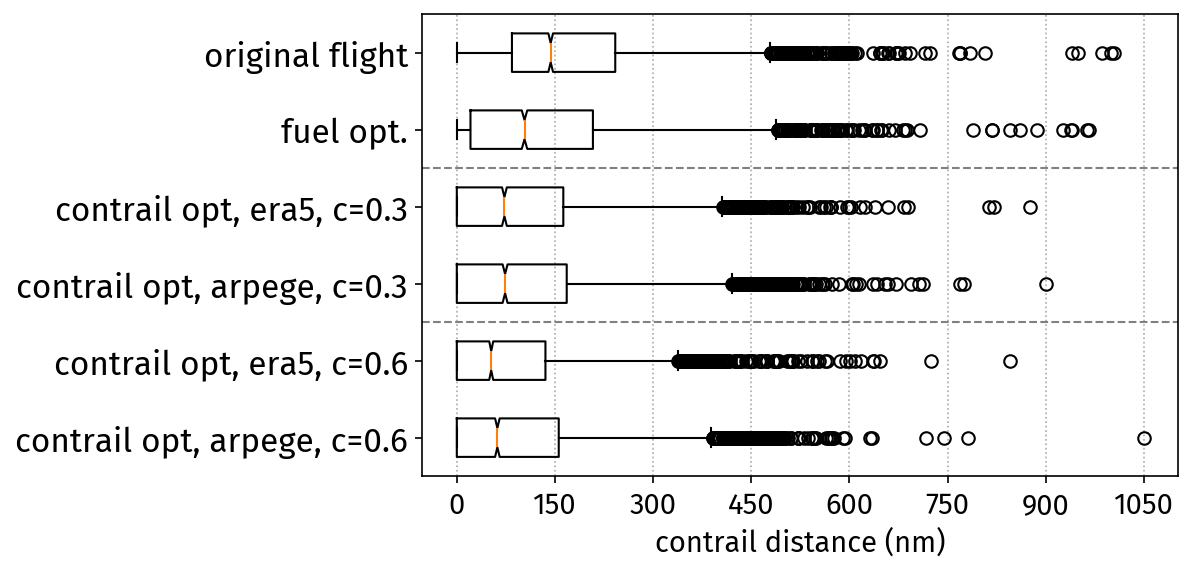}
  \caption{Distribution of contrail distance per flight for contrail-optimal trajectories under different conditions.}
  \label{fig:contrail_distance_boxplot}
\end{figure}

At the network level, we analyzed the contrail distance for all flights in the dataset. Figure \ref{fig:contrail_distance_boxplot} shows the contrail distances for trajectories optimized using both ERA5 reanalysis data and ARPEGE forecast data, where the contrail distances are calculated using the ERA5 data. The boxplot reveals that contrail-optimal trajectories generated using forecast data result in slightly longer contrail distances compared to those using reanalysis data, especially when we want more contrail-aware trajectories ($c=0.6$).

The results from this experiment demonstrate that contrail-aware routing strategies can be sensitive to weather forecast accuracy. For individual flights, discrepancies in contrail formation regions between reanalysis and forecast data can lead to substantially different optimal trajectories. Overall, the total contrail distance can be reduced, but not significant. The forecast uncertainties further limit the effectiveness of contrail mitigation strategies, potentially reducing their expected climate benefits and reducing climate impact during operations.

\section{Are contrail-optimized trajectories sustainable from an air traffic flow management perspective?}
\label{sec:q3}

While most studies focus on reducing contrail impact by adjusting individual flight paths, this approach overlooks broader air traffic management implications. Similar to how not all aircraft can operate at their fuel-optimal altitude due to ATC capacity constraints, widespread adoption of contrail avoidance strategies could significantly strain sector capacities and controller workload.

To assess these network-level impacts, we analyzed airspace capacity at the Area Control Centre (ACC) level over Europe. We calculated traffic density by counting aircraft within each ACC using 10-minute intervals throughout the study day. Then, we compared actual traffic patterns against three optimization scenarios described in Section~\ref{sec:scenario}, which include fuel-optimal trajectories and contrail-optimal trajectories with coefficients of 0.3 and 0.6 compared to fuel, using ARPEGE forecast data.

\begin{figure}[!htbp]
    \centering
    \includegraphics[width=\linewidth]{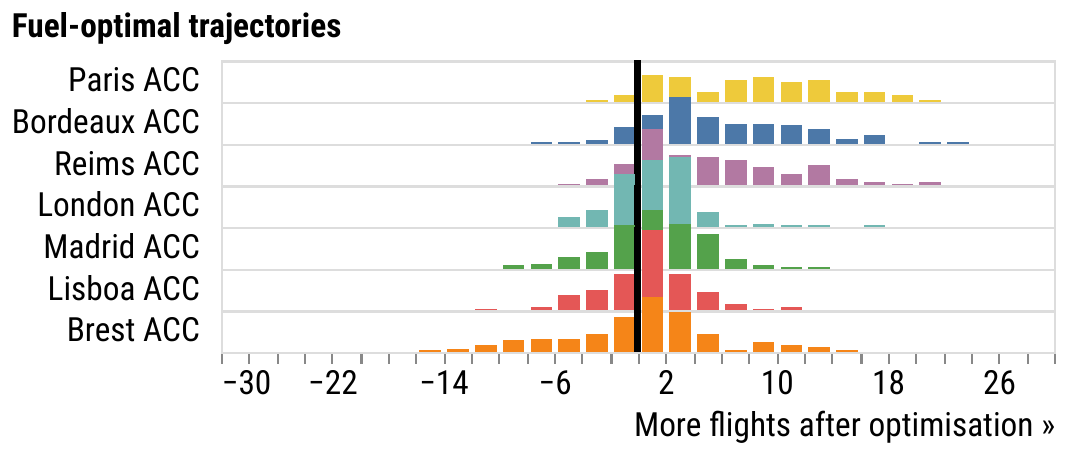}
    \includegraphics[width=\linewidth]{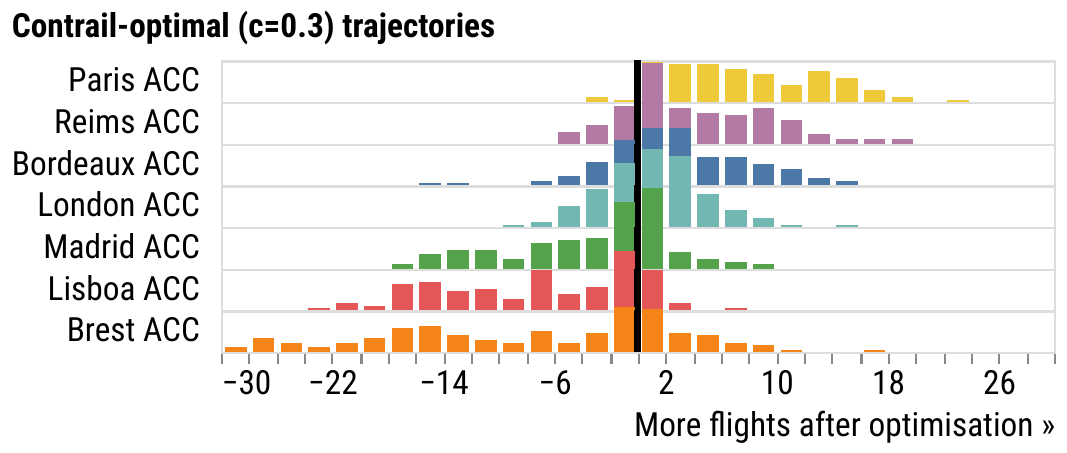}
    \includegraphics[width=\linewidth]{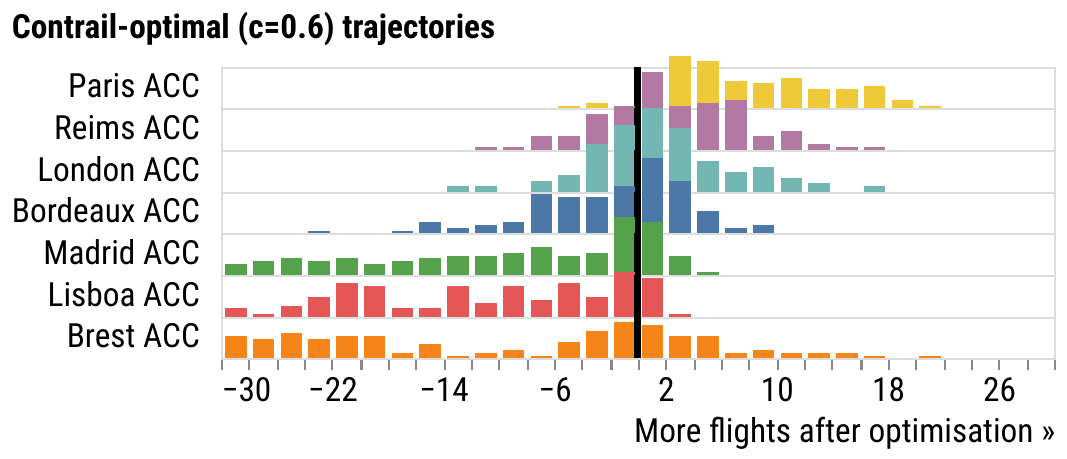}
    \caption{Changes of traffic counts of 10-minute intervals at the ACC level under different trajectory optimization scenarios. $c=0.3$ indicates more fuel-efficient contrail-optimal trajectories, while $c=0.6$ indicates more contrail-aware trajectories, allowing even more excess fuel.}
    \label{fig:capacity}
\end{figure}

Figure~\ref{fig:capacity} presents traffic volume differences, measured in 10-minute intervals, for ACCs showing the highest variability. A rightward shift in distribution indicates increased traffic volume, while a leftward shift shows decreased traffic. Fuel-optimal trajectories generally maintain consistent airspace capacity, with most distributions remaining Gaussian and centered around zero. The main exception is Paris ACC, which handles the traffic at Paris airports. This exception likely reflects the significant fuel inefficiencies in the Paris ACC region due to non-direct routings.

Contrail-optimal trajectories show a different pattern. They tend to reduce traffic in sectors most affected by contrail formation (like Brest, Lisboa, and Madrid ACCs) while increasing controller workload in neighboring sectors such as Bordeaux and Reims ACCs. In extreme cases, some centers experience peaks of up to 15 additional aircraft simultaneously. Such increases could require regulatory measures and cause delays, potentially undermining the intended environmental benefits of contrail avoidance.

Overall, the results suggest that contrail-optimized trajectories may not be sustainable from an air traffic flow management perspective. While they can reduce contrail formation in specific regions, they may lead to capacity issues and increased controller workload in other areas. This trade-off highlights the need for a more systematic approach to contrail mitigation that considers network-wide impacts and operational feasibility.

\section{Who should be responsible for contrail optimal routing mitigation strategies?}
\label{sec:q4}

\subsection{Stakeholders' perspectives}
The implementation of contrail mitigation strategies raises complex political, economic, and operational challenges. Unlike fuel efficiency measures, which offer clear financial incentives, contrail avoidance is not yet supported by a pricing mechanism that would encourage airlines to adopt climate-optimal trajectories voluntarily. As a result, the responsibility for implementing contrail mitigation strategies remains an open question, with possible approaches ranging from airline-level initiatives, ACC-level, FIR-level, or even regional frameworks (e.g., Eurocontrol and FAA)

One of the primary obstacles to an airline-centered market-driven approach is the uncertainty in weather forecasting. Contrail formation is susceptible to atmospheric conditions, and without significant improvements in forecast accuracy, specifically regarding the humidity and clouds, it is difficult to establish a reliable pricing mechanism. Airlines may only be willing to participate in contrail mitigation efforts if a financial incentive or taxation scheme is introduced, yet designing such a system is complicated by the lack of consensus on how to quantify and price the climate impact of contrails.

Additionally, if airlines are expected to take responsibility for reducing their contrails, air traffic control would need to grant them greater flexibility in route planning. However, under current ATC constraints, airlines do not have complete freedom to optimize their trajectories, making it unlikely that they will voluntarily adopt contrail-mitigating strategies. This contradiction raises the question of whether contrail mitigation should instead be managed at the ACC level or at the network level, where traffic flows are already regulated.

\subsection{Stakeholders' monitoring challenges}

Another critical aspect is the ability to monitor contrail formation effectively. Current geostationary satellites, even next-generation systems like Meteosat Third Generation (MTG), lack the spatial resolution needed to accurately track contrails, while low-Earth orbit satellites provide insufficient coverage. Ground-based cameras offer higher resolution but minimal geographic coverage and are ineffective for nighttime contrail monitoring. Without reliable observational data, enforcing contrail mitigation policies could prove difficult.

\begin{figure}[!htbp]
  \centering
  \includegraphics[width=0.9\columnwidth]{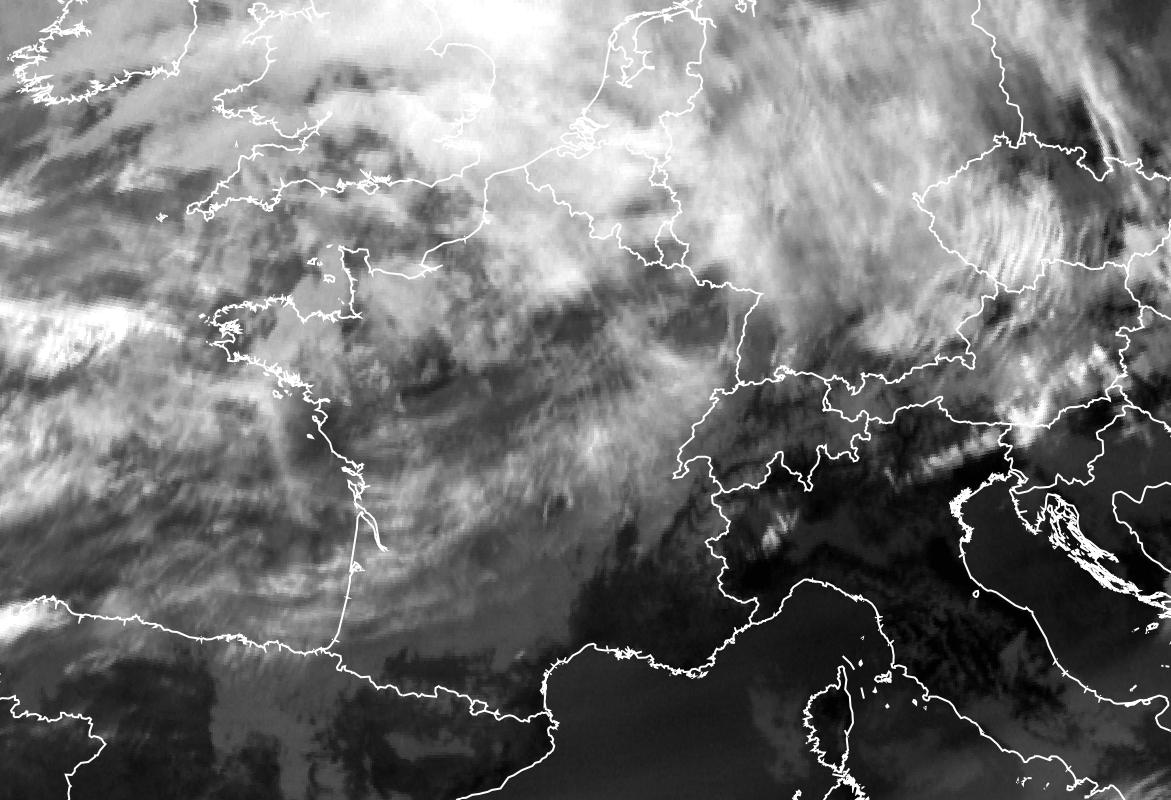}
  \includegraphics[width=0.93\columnwidth]{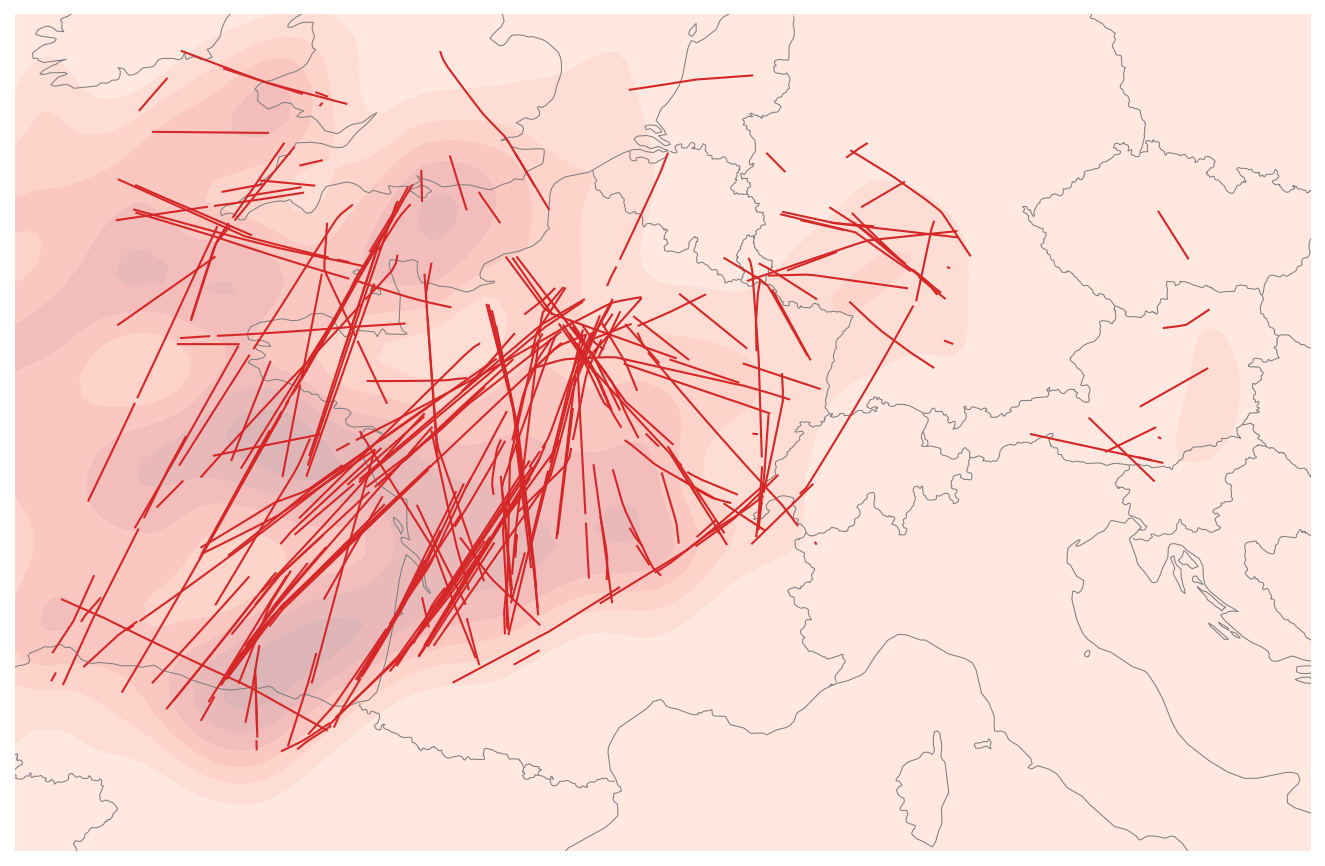}
  \caption{(Top) Meteosat SEVIRI high-rate image data reprojected over part of Western Europe on 20 February 2022, 11:00, provided by EUMETSAT data services. The contrail is hard to see due to natural cloud coverage. (Bottom) Flight trajectories that are predicted to form persistent contrails between 10:00 and 10:45}
  \label{fig:satellite_contrails}
\end{figure}

To examine contrail formation, we analyze satellite imagery. Figure~\ref{fig:satellite_contrails} provides a satellite view of contrails over Western Europe during our study period, along with the persistent contrails estimated from flight data. The top image was captured by SEVIRI (Spinning Enhanced Visible and InfraRed Imager) aboard Meteosat Second Generation satellites and has been reprojected to center on Western Europe. This image is a brightness temperature difference (BTD) product derived from the difference between IR 9.7 and IR 12.0 channels, which highlights contrails and clouds. In the bottom image, we overlay flight trajectories from our dataset on the estimated contrail regions (based on ERA5 data) to compare observed contrails with predicted formation regions.

In the satellite image, both contrails and cirrus clouds appear as white features against a dark background. Linear white features typically indicate aviation-induced contrails, while more diffuse white regions suggest cloud formation. The contrail patterns slightly visible over France align reasonably well with the regions predicted to be conducive to contrail formation based on the Schmidt--Appelmann criterion, as shown earlier in Figures~\ref{fig:era5_cost_grid} and~\ref{fig:optimization_example_era5_vs_arpege}. However, we also observe large areas of clouds where contrails are not visible. 

This comparison highlights several key observations. First, not all aircraft traversing contrail-prone areas appear to generate persistent contrails that can be observed in the satellite image. This comparison suggests that the theoretical predictions based on atmospheric conditions may overestimate actual contrail formation. Second, the image reveals significant cloud coverage in the region. While extensive research has focused on the impact of persistent contrails in clear skies, their interaction with existing cloud systems is not as well understood. 

Contrails forming within or above existing clouds may have different radiative forcing effects compared to those in clear sky conditions due to changes in albedo and atmospheric moisture content. This complexity raises questions about the reliability of current optimization approaches that do not fully account for cloud-contrail interactions.

\subsection{Climate pricing}
A further key challenge in designing a regulatory framework is determining the appropriate climate metric for contrail pricing. For example, should pricing be based on ATR20, ATR50, or ATR100 (which measure climate impact over 20, 50, and 100 years, respectively)? The choice of time horizon significantly affects the perceived severity of contrail-induced climate effects and could lead to disagreements among stakeholders. Establishing a widely accepted standard for contrail impact assessment is crucial for any future regulatory or pricing mechanisms.

The climate impact of contrails is not uniform; compounded contrails, where earlier contrails influence the persistence and radiative effects of subsequent contrails, add another layer of complexity. Research suggests that earlier contrails may have a more significant climate impact than later, overlapping contrails. Should future regulatory frameworks account for this effect when evaluating the climate impact of flights? 

Overall, the question of who should take responsibility for contrail mitigation remains unresolved. While airlines, ANSPs, and regulatory bodies all have a role to play, achieving an effective and equitable solution will require international coordination, improvements in forecasting and monitoring capabilities, and the development of a standardized climate pricing mechanism.

\begin{figure}[!htbp]
  \centering
  \includegraphics[width=\columnwidth]{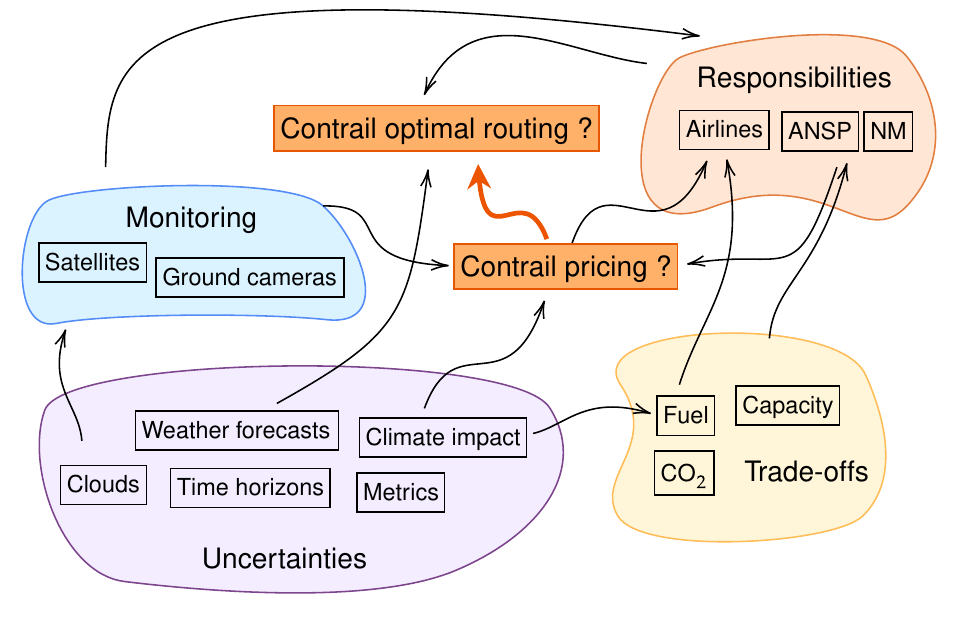}
  \caption{Challenges in implementing operational contrail-optimal routing practices. Contrail pricing is the central missing piece, which remains largely infeasible given current research limitations. The uncertainty in weather forecasts and the lack of monitoring capabilities further complicate the situation. The responsibility for contrail mitigation is still an open question.}
  \label{fig:diagram}
\end{figure}

Finally, the complexity of such a challenge can be shown in the following diagram (Figure \ref{fig:diagram}). Currently, most of the pieces are incomplete in making the contrail optimal routing a reality.

\section*{Conclusion}
\label{sec:conclusion}

In this paper, we considered a day of data over Europe in 2022 and computed some possible state-of-the-art optimizations in flight plans designed to limit the contrail formation on that day. While it is technically within reach to devise aircraft trajectories as solutions to an optimization problem (\ref{sec:q1}), the real impact of such measures is still unclear. We addressed several limitations in terms of uncertainty in the weather forecast (\ref{sec:q2}), but also in the possible effect of those contrails on climate, whether they occur in the clear sky, sum up on top of each other, or appear within other clouds.

Today, airlines play a hide-and-seek game with network managers when they file their flight plans, anticipating possible regulations or possible flight plan changes in order to try to limit their \coo impact as it directly relates to their fuel costs. If airlines were to submit flight plans designed to limit their contrail impact on climate, typical capacity issues would most likely occur (\ref{sec:q3}), leading to a more complex game to resolve a multi-objective optimization problem. The uncertainties in weather forecasts and climate models are all defense mechanisms for airlines to avoid being regulated for their contrail impact on climate. The ethical implications of regulation under uncertainties in aviation are still not well understood, and the question of who should be responsible for contrail mitigation remains open (\ref{sec:q4}).

Lastly, the impact of contrail formation is at present still not economically quantifiable (\ref{sec:q4}), and there is no incentive for airlines or airspace operators to limit their contrail impact on climate. Furthermore, if any stakeholder were to be charged for the contrail they produce, technology would be behind in terms of capabilities to monitor any contrail formation and match it to a given aircraft.

Limiting aviation's impact on climate remains a critical challenge. While contrail mitigation strategies have been extensively studied in academia, their implementation in real-world operations is still in its early stages. We now have access to high-quality trajectory and weather data, mature mathematical optimization tools, and sufficient computational power to conduct robust simulations. However, even then, the uncertainties associated with the climate impact of contrails still pose significant issues in the adoption of actual operations. One must question whether the current state of knowledge is sufficient to justify the contrail-aware routing strategies in practice.

However, beyond technical advancements, progress also depends on political decisions, regulatory frameworks, and scientists to refine climate models and better understand contrail dynamics. Perhaps, as seen in other sectors like energy, the immediate priority should be achieving zero \coo emissions first, with other climate forcers, such as contrails, addressed in a subsequent phase when contrail's impact on climate can be scientifically quantified.

\section*{Reproducibility statement}

All data and code used in this research are openly available to ensure reproducibility and transparency. The code and data support this research can be found at:

\begin{center}
\fbox{
  \parbox{0.9\linewidth}{
    \centering
    \url{https://github.com/junzis/contrail-or-not}
  }
}
\end{center}

\vspace{1em}

\bibliographystyle{IEEEtran}
\bibliography{reference.bib}

\end{document}